\documentclass{article}
\usepackage{graphicx}
\usepackage{amsmath}

\input{tcilatex}

\begin{document}

\title{Generalized Quantization Scheme for Two-Person Non-Zero-Sum Games}
\author{Ahmad Nawaz\thanks{%
ahmad@ele.qau.edu.pk} and A. H. Toor\thanks{%
ahtoor@qau.edu.pk} \\
Department of Electronics, Quaid-i-Azam University, \ \ \ \ \ \\
Islamabad 45320, Pakistan.}
\maketitle

\begin{abstract}
We have proposed a generalized quantization scheme for non-zero sum games
which can be reduced to two existing quantization schemes under appropriate
set of parameters. Some other important situations are identified which are
not apparent in the exiting two quantizations schemes.
\end{abstract}

\section{Introduction}

Game theory stepped into quantum domain with the success of a hypothetical
quantum player over a classical player in a Quantum\emph{\ }Penny Flip game %
\cite{meyer,Furtherreading}.\emph{\ }Later Eisert et. al. \cite{eisert}
introduced an elegant scheme to deal with non-zero sum games quantum
mechanically. In this quantization scheme the strategy space of the players
is a two parameter set of unitary $2\times 2$ matrices. Starting with
maximally entangled initial state they analyzed a well-known Prisoner
Dilemma game and showed that for a suitable quantum strategy the dilemma
disappears. They also pointed out a quantum strategy which always wins over
all the classical strategies. Later on, Marinatto and Weber \cite{marin}
introduced another interesting and simple scheme for the analysis of
non-zero sum games in quantum domain. They gave Hilbert structure to the
strategic spaces of the players. They used maximally entangled initial state
and allowed the players to play their tactics by applying the probabilistic
choice of unitary operators. They applied their scheme to an interesting
game of Battle of Sexes and found out the strategy for which both the
players can achieve equal payoffs.

Both Eisert's and Marinatto and Weber's\emph{\ }schemes give interesting
results for various quantum versions of the games \cite%
{azhar,azhar1,jiang1,flitney,Rosero,nawaz}. It seems natural to look for a
relationship between these two apparently different quantization schemes. In
this papers we have developed a generalized quantization scheme for non zero
sum games. The game of Battle of Sexes has been used as an example to
introduce this quantization scheme which is applicable to other games as
well. Separate set of parameters are identified for which our scheme reduces
to that of Marinatto and Weber \cite{marin} and Eisert et al \cite{eisert}
schemes. Furthermore we have identified other interesting situations which
are not apparent within the exiting two quantizations schemes. After a brief
introduction to Battle of Sexes we have extended Marinatto and Weber's
mathematical framework by redefining unitary operators for our generalized
quantization scheme.

\section{\label{battle} Generalized Quantization Scheme}

Battle of sexes is an interesting static game of complete information. In
its usual exposition two players Alice and Bob are trying to decide a place
to spend Saturday evening. Alice wants to attend Opera while Bob is
interested in watching TV at home and both would prefer to spend the evening
together. The game is represented by the following payoff matrix:

\ \ \ \ \ \ \ \ \ \ \ \ \ \ \ \ \ \ \ \ \ \ \ \ \ \ \ \ \ \ \ \ \ \ \ \ \ \
\ \ \ Bob

$\ \ \ \ \ \ \ \ \ \ \ \ \ \ \ \ \ \ \ \ \ \ \ \ \ \ \ \ 
\begin{array}{c}
\end{array}
\ \ \ \ \ 
\begin{array}{cc}
\text{ }O\text{ \ \ \ \ \ \ } & T%
\end{array}
$

\ \ \ \ \ \ \ \ \ \ \ \ \ \ \ \ \ \ \ Alice$ 
\begin{array}{c}
O \\ 
T%
\end{array}
\left[ 
\begin{array}{cc}
(\alpha ,\beta ) & (\sigma ,\sigma ) \\ 
(\sigma ,\sigma ) & (\beta ,\alpha )%
\end{array}
\right] ,$

where $O$ and $T$ represent Opera and TV, respectively, and $\alpha $, $%
\beta $, $\sigma $ are the payoffs for players for different choices of
strategies with $\alpha >\beta >\sigma $. There are two Nash equilibria $%
(O,O)$ and $(T,T)$ existing in the classical form of the game. In absence of
any communication between Alice and Bob, there exists a dilemma as Nash
equilibria $\left( O,O\right) $ suits Alice whereas Bob prefers $(T,T).$ As
a result both the players could end up with worst payoff in the case they
play mismatched strategies. Marinatto and Weber \cite{marin} presented the
quantum version of the game to resolve this dilemma. In our earlier paper we
have further extended their work to remove the worst case payoff situation
in Battle of Sexes\cite{nawaz}. On the other hand Eisert et. al. \cite%
{eisert} presented a different scheme to remove dilemma in the game of
Prisoner's Dilemma through quantization of the game.

Here we present generalized quantization scheme by redefining unitary
operators in the Marinatto and Weber scheme. Let Alice and Bob are given the
following initial state 
\begin{equation}
\left| \psi _{in}\right\rangle =\cos \frac{\gamma }{2}\left| OO\right\rangle
+i\sin \frac{\gamma }{2}\left| TT\right\rangle .  \label{state in}
\end{equation}
Here $\left| O\right\rangle $ and $\left| T\right\rangle $ represent the
vectors in the strategy space corresponding to Opera and TV, respectively%
\emph{\ }and\emph{\ }$\gamma \in \left[ 0,\frac{\pi }{2}\right] $.\emph{\ }%
The strategy of each of the\ players is represented by the unitary operator $%
U_{i}$\ of the form\emph{\ } 
\begin{equation}
U_{i}=\cos \frac{\theta _{i}}{2}R_{i}+\sin \frac{\theta _{i}}{2}C_{i},\text{
\ \ \ }  \label{combination}
\end{equation}
where $i=1$\ or $2$\ and $R$, $C$\emph{\ }are the unitary operators defined
as:

\begin{align}
R\left| O\right\rangle & =e^{i\phi _{i}}\left| O\right\rangle ,\text{ \ \ }%
R\left| T\right\rangle =e^{-i\phi _{i}}\left| T\right\rangle ,  \notag \\
C\left| O\right\rangle & =-\left| T\right\rangle ,\text{ \ \ \ \ \ }C\left|
T\right\rangle =\left| O\right\rangle .  \label{oper}
\end{align}%
Here we restrict our treatment to two parameter set of strategies for
mathematical simplicity in accordance with Ref. \cite{eisert}.\emph{\ }After
the application of the strategies, the initial state (\ref{state in})
transforms into 
\begin{equation}
\left| \psi _{f}\right\rangle =(U_{1}\otimes U_{2})\left| \psi
_{in}\right\rangle .  \label{final}
\end{equation}%
and using eqs. (\ref{combination}) and (\ref{oper}) the above expression
becomes- 
\begin{align}
\left| \psi _{f}\right\rangle & =\cos \frac{\gamma }{2}[\cos \frac{\theta
_{1}}{2}\cos \frac{\theta _{2}}{2}e^{i(\phi _{1}+\phi _{2})}\left|
OO\right\rangle -\cos \frac{\theta _{1}}{2}\sin \frac{\theta _{2}}{2}%
e^{i\phi _{1}}\left| OT\right\rangle   \notag \\
& -\cos \frac{\theta _{2}}{2}\sin \frac{\theta _{1}}{2}e^{i\phi _{2}}\left|
TO\right\rangle +\sin \frac{\theta _{1}}{2}\sin \frac{\theta _{2}}{2}\left|
TT\right\rangle ]  \notag \\
& +i\sin \frac{\gamma }{2}[\cos \frac{\theta _{1}}{2}\cos \frac{\theta _{2}}{%
2}e^{-i(\phi _{1}+\phi _{2})}\left| TT\right\rangle +\cos \frac{\theta _{1}}{%
2}\sin \frac{\theta _{2}}{2}e^{-i\phi _{1}}\left| TO\right\rangle   \notag \\
& +\cos \frac{\theta _{2}}{2}\sin \frac{\theta _{1}}{2}e^{-i\phi _{2}}\left|
OT\right\rangle +\sin \frac{\theta _{1}}{2}\sin \frac{\theta _{2}}{2}\left|
OO\right\rangle ].  \label{state fin}
\end{align}%
The payoff operators for Alice and Bob are

\begin{align}
P_{A}& =\alpha P_{OO}+\beta P_{TT}+\sigma (P_{OT}+P_{TO})  \notag \\
P_{B}& =\alpha P_{TT}+\beta P_{OO}+\sigma (P_{OT}+P_{TO})
\label{pay-operator}
\end{align}%
where 
\begin{subequations}
\label{oper a}
\begin{align}
P_{OO}& =\left| \psi _{OO}\right\rangle \left\langle \psi _{OO}\right| \text{%
, \ }\left| \psi _{OO}\right\rangle =\cos \frac{\delta }{2}\left|
OO\right\rangle +i\sin \frac{\delta }{2}\left| TT\right\rangle ,
\label{oper 1} \\
P_{TT}& =\left| \psi _{TT}\right\rangle \left\langle \psi _{TT}\right| ,%
\text{ \ }\left| \psi _{TT}\right\rangle =\cos \frac{\delta }{2}\left|
TT\right\rangle +i\sin \frac{\delta }{2}\left| OO\right\rangle ,
\label{oper 2} \\
P_{TO}& =\left| \psi _{TO}\right\rangle \left\langle \psi _{TO}\right| \text{%
, \ }\left| \psi _{TO}\right\rangle =\cos \frac{\delta }{2}\left|
TO\right\rangle -i\sin \frac{\delta }{2}\left| OT\right\rangle ,
\label{oper 3} \\
P_{OT}& =\left| \psi _{OT}\right\rangle \left\langle \psi _{OT}\right| \text{%
, \ }\left| \psi _{OT}\right\rangle =\cos \frac{\delta }{2}\left|
OT\right\rangle -i\sin \frac{\delta }{2}\left| TO\right\rangle .
\label{oper 4}
\end{align}%
with\emph{\ }$\delta \in \left[ 0,\frac{\pi }{2}\right] $.Above payoff
operators reduce to that of Eisert's scheme for $\delta $ equal to $\gamma ,$
which represents the entanglement of the initial state. And for $\delta =0$
above operators transform into that of Marinatto and Weber's scheme. In
generalized quantization scheme payoff for the players are calculated as %
\cite{footnote} 
\end{subequations}
\begin{eqnarray}
\$_{A}(\theta _{1},\phi _{1},\theta _{2},\phi _{2}) &=&\text{Tr}(P_{A}\rho
_{f})\text{,}  \notag \\
\$_{B}(\theta _{1},\phi _{1},\theta _{2},\phi _{2}) &=&\text{Tr}(P_{B}\rho
_{f}),  \label{payoff}
\end{eqnarray}%
where $\rho _{f}=\left| \psi _{f}\right\rangle \left\langle \psi _{f}\right| 
$ is the density matrix for the quantum state given by (\ref{state fin}) and
Tr represents the trace of a\emph{\ }matrix. Using eqs. (\ref{state fin}, %
\ref{pay-operator}, \ref{payoff}) the payoffs for players are obtained as 
\begin{subequations}
\label{oper a}
\begin{align}
\$_{A}(\theta _{1},\phi _{1},\theta _{2},\phi _{2})& =\cos ^{2}\frac{\theta
_{1}}{2}\cos ^{2}\frac{\theta _{2}}{2}\left[ \eta \sin ^{2}\frac{\gamma }{2}%
+\xi \cos ^{2}\frac{\gamma }{2}+\chi \cos 2(\phi _{1}+\phi _{2})\sin \gamma
\right.  \notag \\
& \left. -\sigma \right] +\sin ^{2}\frac{\theta _{1}}{2}\sin ^{2}\frac{%
\theta _{2}}{2}(\eta \cos ^{2}\frac{\gamma }{2}+\xi \sin ^{2}\frac{\gamma }{2%
}-\chi \sin \gamma -\sigma )  \notag \\
& +\frac{(\alpha +\beta -2\sigma )\sin \gamma -2\chi }{4}\sin \theta
_{1}\sin \theta _{2}\sin (\phi _{1}+\phi _{2})+\sigma  \notag \\
&  \label{GPA} \\
\$_{B}(\theta _{1},\phi _{1},\theta _{2},\phi _{2})& =\cos ^{2}\frac{\theta
_{1}}{2}\cos ^{2}\frac{\theta _{2}}{2}\left[ \xi \sin ^{2}\frac{\gamma }{2}%
+\eta \cos ^{2}\frac{\gamma }{2}-\chi \cos 2(\phi _{1}+\phi _{2})\sin \gamma
\right.  \notag \\
& \left. -\sigma \right] +\sin ^{2}\frac{\theta _{1}}{2}\sin ^{2}\frac{%
\theta _{2}}{2}(\xi \cos ^{2}\frac{\gamma }{2}+\eta \sin ^{2}\frac{\gamma }{2%
}+\chi \sin \gamma -\sigma )+  \notag \\
& \frac{(\alpha +\beta -2\sigma )\sin \gamma +2\chi }{4}\sin \theta _{1}\sin
\theta _{2}\sin (\phi _{1}+\phi _{2})+\sigma ,  \notag \\
&  \label{GPB}
\end{align}%
where $\xi =\alpha \cos ^{2}\frac{\delta }{2}+\beta \sin ^{2}\frac{\delta }{2%
},$ $\eta =\alpha \sin ^{2}\frac{\delta }{2}+\beta \cos ^{2}\frac{\delta }{2}%
,$ and $\chi =\frac{(\alpha -\beta )}{2}\sin \delta $. Classical results can
easily be found from eqs (\ref{GPA},\ref{GPB}) by simply unentangling, the
initial quantum state of the game i.e. letting $\gamma =0$\emph{.}
Furthermore all the results found by Marinatto and Weber \cite{marin} and
Eisert et. al. \cite{eisert} are also embedded in these payoffs. For
different combinations of $\delta $ and $\phi ^{\prime }s$ there arise the
following possibilities

\textbf{Case(a): }When $\delta =0$ and

(i) $\phi _{1}=0$, $\phi _{2}=0.$ then the payoffs for the players from eqs (%
\ref{GPA},\ref{GPB}) reduce to 
\end{subequations}
\begin{subequations}
\label{marinto}
\begin{align}
\$_{A}(\theta _{1},\phi _{1,}\theta _{2},\phi _{2})& =\cos ^{2}\frac{\theta
_{1}}{2}[\cos ^{2}\frac{\theta _{2}}{2}(\alpha +\beta -2\sigma )-\alpha \sin
^{2}\frac{\gamma }{2}-\beta \cos ^{2}\frac{\gamma }{2}+\sigma ]  \notag \\
& +\cos ^{2}\frac{\theta _{2}}{2}(-\alpha \sin ^{2}\frac{\gamma }{2}-\beta
\cos ^{2}\frac{\gamma }{2}+\sigma )+\alpha \sin ^{2}\frac{\gamma }{2}+\beta
\cos ^{2}\frac{\gamma }{2}  \label{marintoa} \\
\$_{B}(\theta _{1},\phi _{1},\theta _{2},\phi _{2})& =\cos ^{2}\frac{\theta
_{2}}{2}[\cos ^{2}\frac{\theta _{1}}{2}(\alpha +\beta -2\sigma )-\beta \sin
^{2}\frac{\gamma }{2}-\alpha \cos ^{2}\frac{\gamma }{2}+\sigma ]  \notag \\
& +\cos ^{2}\frac{\theta _{1}}{2}(-\beta \sin ^{2}\frac{\gamma }{2}-\alpha
\cos ^{2}\frac{\gamma }{2}+\sigma )+\beta \sin ^{2}\frac{\gamma }{2}+\alpha
\cos ^{2}\frac{\gamma }{2}  \label{marinatob}
\end{align}
These payoffs are the same as found by Marinatto and Weber \cite{marin} when
the players are applying the identity operators $I_{1}$ and $I_{2}$ with
probabilities $\cos ^{2}\frac{\theta _{1}}{2}$ and $\cos ^{2}\frac{\theta
_{2}}{2}$ respectively for the given initial quantum state of the form (\ref%
{state in}).

(ii) $\phi _{1}+\phi _{2}=\frac{\pi }{2}$ eqs (\ref{GPA},\ref{GPB}) reduce to

\end{subequations}
\begin{subequations}
\label{marin+chinese1}
\begin{align}
\$_{A}(\theta _{1},\phi _{1},\theta _{2},\phi _{2})& =\cos ^{2}\frac{\theta
_{1}}{2}\left[ \cos ^{2}\frac{\theta _{2}}{2}(\alpha +\beta -2\sigma
)-\alpha \sin ^{2}\frac{\gamma }{2}-\beta \cos ^{2}\frac{\gamma }{2}+\sigma %
\right]  \notag \\
& +\cos ^{2}\frac{\theta _{2}}{2}\left( -\alpha \sin ^{2}\frac{\gamma }{2}%
-\beta \cos ^{2}\frac{\gamma }{2}+\sigma \right) +\alpha \sin ^{2}\frac{%
\gamma }{2}+\beta \cos ^{2}\frac{\gamma }{2}  \notag \\
& +\frac{\left( \alpha +\beta -2\sigma \right) }{4}\sin \gamma \sin \theta
_{1}\sin \theta _{2}  \label{marin+chines1a} \\
\$_{B}(\theta _{1},\phi _{1},\theta _{2},\phi _{2})& =\cos ^{2}\frac{\theta
_{2}}{2}\left[ \cos ^{2}\frac{\theta _{1}}{2}(\alpha +\beta -2\sigma )-\beta
\sin ^{2}\frac{\gamma }{2}-\alpha \cos ^{2}\frac{\gamma }{2}+\sigma \right] 
\notag \\
& +\cos ^{2}\frac{\theta _{1}}{2}\left( -\beta \sin ^{2}\frac{\gamma }{2}%
-\alpha \cos ^{2}\frac{\gamma }{2}+\sigma \right) +\beta \sin ^{2}\frac{%
\gamma }{2}+\alpha \cos ^{2}\frac{\gamma }{2}  \notag \\
& +\frac{\left( \alpha +\beta -2\sigma \right) }{4}\sin \gamma \sin \theta
_{1}\sin \theta _{2}  \label{marin+chinese1a}
\end{align}
These payoffs are equivalent to as if the players are using a linear
combination of operators $I$ and flip operator $\sigma _{x}$ of the form $%
O_{i}=\sqrt{p_{i}}I+\sqrt{1-p_{i}}\sigma _{x}$ where $p_{i}=\cos ^{2}\frac{%
\theta _{i}}{2}$, $i=1$ or $2$ using Marinatto and Weber scheme \cite%
{marin,Ying}, for the initial entangled state of the form of eq. (\ref{state
in}).

\textbf{Case (b)} When $\delta =\gamma $ and

(i)$\ \phi _{1}\neq 0$, $\phi _{2}\neq 0$ the payoffs given by the eqs (\ref%
{GPA},\ref{GPB}) very interestingly change to the payoffs as if the game has
been quantized using Eisert et. al.\cite{eisert} scheme for the initial
quantum state of the form (\ref{state in}). In this situation the payoffs
for both the players are 
\end{subequations}
\begin{subequations}
\label{marin+chinese1}
\begin{gather}
\$_{A}(\theta _{1},\phi _{1},\theta _{2},\phi _{2})=\cos ^{2}\frac{\theta
_{1}}{2}\cos ^{2}\frac{\theta _{2}}{2}\left[ \eta _{1}\sin ^{2}\frac{\gamma 
}{2}+\xi _{1}\cos ^{2}\frac{\gamma }{2}+\chi _{1}\cos 2(\phi _{1}+\phi
_{2})\right.  \notag \\
\left. -\sigma \right] +\sin ^{2}\frac{\theta _{1}}{2}\sin ^{2}\frac{\theta
_{2}}{2}\left( \eta _{1}\cos ^{2}\frac{\gamma }{2}+\xi _{1}\sin ^{2}\frac{%
\gamma }{2}-\chi _{1}-\sigma \right)  \notag \\
+\frac{\left( \beta -\sigma \right) }{2}\sin \gamma \sin \theta _{1}\sin
\theta _{2}\sin \left( \phi _{1}+\phi _{2}\right) +\sigma
\label{payoff-general1} \\
\$_{B}(\theta _{1},\phi _{1},\theta _{2},\phi _{2})=\cos ^{2}\frac{\theta
_{1}}{2}\cos ^{2}\frac{\theta _{2}}{2}\left[ \xi _{1}\sin ^{2}\frac{\gamma }{%
2}+\eta _{1}\cos ^{2}\frac{\gamma }{2}-\chi _{1}\cos 2\left( \phi _{1}+\phi
_{2}\right) \right.  \notag \\
\left. -\sigma \right] +\sin ^{2}\frac{\theta _{1}}{2}\sin ^{2}\frac{\theta
_{2}}{2}\left( \xi _{1}\cos ^{2}\frac{\gamma }{2}+\eta _{1}\sin ^{2}\frac{%
\gamma }{2}+\chi _{1}-\sigma \right)  \notag \\
+\frac{\left( \alpha -\sigma \right) }{2}\sin \gamma \sin \theta _{1}\sin
\theta _{2}\sin \left( \phi _{1}+\phi _{2}\right) +\sigma
\label{payoff-general2}
\end{gather}
where $\xi _{1}=\alpha \cos ^{2}\frac{\gamma }{2}+\beta \sin ^{2}\frac{%
\gamma }{2},$ $\eta _{1}=\alpha \sin ^{2}\frac{\gamma }{2}+\beta \cos ^{2}%
\frac{\gamma }{2},$ and $\chi _{1}=\frac{(\alpha -\beta )}{2}\sin ^{2}\gamma 
$. To draw a better comparison we take $\delta =\gamma =\frac{\pi }{2}$ then
the payoffs given by eqs (\ref{marin+chinese1}) reduce to 
\end{subequations}
\begin{subequations}
\label{payoffs}
\begin{align}
\$_{A}(\theta _{1},\phi _{1},\theta _{2},\phi _{2})& =\left( \alpha -\sigma
\right) \cos ^{2}\frac{\theta _{1}}{2}\cos ^{2}\frac{\theta _{2}}{2}\sin
^{2}\left( \phi _{1}+\phi _{2}\right)  \notag \\
& +\left( \beta -\sigma \right) \left[ \cos \frac{\theta _{1}}{2}\cos \frac{%
\theta _{2}}{2}\sin (\phi _{1}+\phi _{2})+\sin \frac{\theta _{1}}{2}\sin 
\frac{\theta _{2}}{2}\right] ^{2}+\sigma  \label{payoff-A1} \\
\$_{B}(\theta _{1},\phi _{1},\theta _{2},\phi _{2})& =\left( \alpha -\sigma
\right) \left[ \cos \frac{\theta _{1}}{2}\cos \frac{\theta _{2}}{2}\sin
(\phi _{1}+\phi _{2})+\sin \frac{\theta _{1}}{2}\sin \frac{\theta _{2}}{2}%
\right] ^{2}  \notag \\
& +\left( \beta -\sigma \right) \cos ^{2}\frac{\theta _{1}}{2}\cos ^{2}\frac{%
\theta _{2}}{2}\sin ^{2}\left( \phi _{1}+\phi _{2}\right) +\sigma
\label{payoff-B1}
\end{align}
The payoffs given in eqs (\ref{payoffs})\ have already been found by J. Du
et. al. \cite{chinese} through Eisert et. al. scheme \cite{eisert}.

(ii) $\phi _{1}=\phi _{2}=0$ As shown by Eisert et. al. \cite{eisert,eisert1}
that one gets classical payoffs with mixed strategies when one parameter set
of strategies\ is used for the quantization of the game. For a better
comparison putting $\gamma =\delta =\frac{\pi }{2}$ and $\phi _{1}=\phi
_{2}=0$ in eqs (\ref{payoff-general1}) and (\ref{payoff-general2}) the same
situation occurs and the payoffs reduce to 
\end{subequations}
\begin{subequations}
\label{one-parameter}
\begin{eqnarray}
\$_{A}(\theta _{1},\phi _{1},\theta _{2},\phi _{2}) &=&\alpha \cos ^{2}\frac{%
\theta _{1}}{2}\cos ^{2}\frac{\theta _{2}}{2}+\beta \sin ^{2}\frac{\theta
_{1}}{2}\sin ^{2}\frac{\theta _{2}}{2}  \notag \\
&&+\sigma (\cos ^{2}\frac{\theta _{1}}{2}\sin ^{2}\frac{\theta _{2}}{2}+\sin
^{2}\frac{\theta _{1}}{2}\cos ^{2}\frac{\theta _{2}}{2})
\label{one-parametera} \\
\$_{B}(\theta _{1},\phi _{1},\theta _{2},\phi _{2}) &=&\beta \cos ^{2}\frac{%
\theta _{1}}{2}\cos ^{2}\frac{\theta _{2}}{2}+\alpha \sin ^{2}\frac{\theta
_{1}}{2}\sin ^{2}\frac{\theta _{2}}{2}  \notag \\
&&+\sigma (\cos ^{2}\frac{\theta _{1}}{2}\sin ^{2}\frac{\theta _{2}}{2}+\sin
^{2}\frac{\theta _{1}}{2}\cos ^{2}\frac{\theta _{2}}{2})
\label{one-parameterb}
\end{eqnarray}
In this case the game behaves just like classical game where the players are
playing mixed strategies with probabilities $\cos ^{2}\frac{\theta _{1}}{2}$
and $\cos ^{2}\frac{\theta _{2}}{2}$ respectively.

\textbf{Case (c}) when $\delta \neq \gamma $\ and $\phi _{1}=0$, $\phi
_{2}=0 $\ the payoffs given by the eqs (\ref{GPA},\ref{GPB}) reduce to 
\end{subequations}
\begin{subequations}
\label{one-parameter}
\begin{gather}
\$_{A}(\theta _{1},\phi _{1,}\theta _{2},\phi _{2}=\cos ^{2}\frac{\theta _{1}%
}{2}\left[ \cos ^{2}\frac{\theta _{2}}{2}\left( \alpha +\beta -2\sigma
\right) -\alpha \sin ^{2}\frac{\left( \gamma -\delta \right) }{2}\right. 
\notag \\
\left. -\beta \cos ^{2}\frac{(\gamma -\delta )}{2}+\sigma \right] +\cos ^{2}%
\frac{\theta _{2}}{2}\left[ -\alpha \sin ^{2}\frac{\left( \gamma -\delta
\right) }{2}\right.  \notag \\
\left. -\beta \cos ^{2}\frac{\left( \gamma -\delta \right) }{2}+\sigma %
\right] +\alpha \sin ^{2}\frac{\left( \gamma -\delta \right) }{2}+\beta \cos
^{2}\frac{\left( \gamma -\delta \right) }{2}  \label{delta0} \\
\$_{B}\left( \theta _{1},\phi _{1,}\theta _{2},\phi _{2}\right) =\cos ^{2}%
\frac{\theta _{1}}{2}\left[ \cos ^{2}\frac{\theta _{2}}{2}\left( \alpha
+\beta -2\sigma \right) -\beta \sin ^{2}\frac{\left( \gamma -\delta \right) 
}{2}\right.  \notag \\
\left. -\alpha \cos ^{2}\frac{\left( \gamma -\delta \right) }{2}+\sigma %
\right] +\cos ^{2}\frac{\theta _{2}}{2}\left[ -\beta \sin ^{2}\frac{\left(
\gamma -\delta \right) }{2}\right.  \notag \\
\left. -\alpha \cos ^{2}\frac{\left( \gamma -\delta \right) }{2}+\sigma %
\right] +\beta \sin ^{2}\frac{\left( \gamma -\delta \right) }{2}+\alpha \cos
^{2}\frac{\left( \gamma -\delta \right) }{2}  \label{delta}
\end{gather}

These payoffs are equivalent to Marinatto and Weber \cite{marin} when $%
\gamma $ replaced with $\gamma -\delta .$

\textbf{Case (d) }When $\delta \neq 0$ and $\gamma =0$ then from eqs (\ref%
{GPA},\ref{GPB}) the payoffs of the players reduce to 
\end{subequations}
\begin{subequations}
\label{one-parameter}
\begin{align}
\$_{A}(\theta _{1},\phi _{1},\phi _{2},\theta _{2})& =\cos ^{2}\frac{\theta
_{1}}{2}\left[ \cos ^{2}\frac{\theta _{2}}{2}\left( \alpha +\beta -2\sigma
\right) -\alpha \sin ^{2}\frac{\delta }{2}-\beta \cos ^{2}\frac{\delta }{2}%
+\sigma \right]  \notag \\
& +\cos ^{2}\frac{\theta _{2}}{2}\left( -\alpha \sin ^{2}\frac{\delta }{2}%
-\beta \cos ^{2}\frac{\delta }{2}+\sigma \right) +\alpha \sin ^{2}\frac{%
\delta }{2}+\beta \cos ^{2}\frac{\delta }{2}  \notag \\
& -\frac{\left( \alpha -\beta \right) }{2}\sin \delta \sin \theta _{1}\sin
\theta _{2}\sin \left( \phi _{1}+\phi _{2}\right) \\
\$_{B}(\theta _{1},\phi _{1},\phi _{2},\theta _{2})& =\cos ^{2}\frac{\theta
_{2}}{2}\left[ \cos ^{2}\frac{\theta _{1}}{2}\left( \alpha +\beta -2\sigma
\right) -\beta \sin ^{2}\frac{\delta }{2}-\alpha \cos ^{2}\frac{\delta }{2}%
+\sigma \right]  \notag \\
& +\cos ^{2}\frac{\theta _{1}}{2}\left( -\beta \sin ^{2}\frac{\delta }{2}%
-\alpha \cos ^{2}\frac{\delta }{2}+\sigma \right) +\beta \sin ^{2}\frac{%
\delta }{2}+\alpha \cos ^{2}\frac{\delta }{2}  \notag \\
& +\frac{\left( \alpha -\beta \right) }{2}\sin \delta \sin \theta _{1}\sin
\theta _{2}\sin \left( \phi _{1}+\phi _{2}\right)  \label{un-entangled}
\end{align}
This shows that the measurement plays a crucial role in quantum games as if
initial state is unentangled, i.e., $\gamma =0,$arbiter can still apply
entangled basis for the measurement to obtain quantum mechanical results.
Above payoff's are similar to that of Marinatto and Weber for the Battle of
Sexes games if $\delta $ is replaced by $\gamma .$

\section{Conclusion}

A generalized quantization scheme for non zero sum games is proposed. The
game of Battle of Sexes has been used as an example to introduce this
quantization scheme. However our quantization scheme is applicable to other
games as well. This new scheme reduces to Eisert's et al \cite{eisert}
scheme under the condition 
\end{subequations}
\begin{equation*}
\delta =\gamma ,\phi _{1}+\phi _{2}=\pi /2
\end{equation*}
and to Marinatto and Weber \cite{marin} scheme when 
\begin{equation*}
\delta =0,\phi _{1}=0,\phi _{2}=0.
\end{equation*}
In the above conditions $\gamma $ is a measure of entanglement of the
initial state. For $\gamma =0,$ classical results are obtained when $\delta
=0,\phi _{1}=0,\phi _{2}=0$. Furthermore, we have identified some
interesting situations which are not apparent within the exiting two
quantizations schemes. For example, with $\delta \neq 0,$\ nonclassical
results are obtained for initially unentangled state. This shows that the
measurement plays a crucial role in quantum games.

\end{document}